\documentclass[epj]{webofc}
\usepackage[varg]{txfonts} 
\usepackage{float} 
\wocname{EPJ Web of Conferences}
\woctitle{CONF12}
\begin{document}
\selectlanguage{english}
\title{Direct-Photon Spectra and Anisotropic Flow in Heavy Ion Collisions from Holography}
%
%

\author{Ioannis Iatrakis\inst{1}\and
        Elias Kiritsis\inst{2,3,4} \and
        Chun Shen\inst{5} \and 
        Di-Lun Yang\inst{6}\fnsep\thanks{\email{dilunyang@gmail.com}}\
}

\institute{Institute for Theoretical Physics and Center for Extreme Matter and Emergent Phenomena, Utrecht University, Leuvenlaan 4, 3584 CE Utrecht, The Netherlands.
\and
           Crete Center for Theoretical Physics\text{,} Institute of Theoretical and Computational Physics, Department of Physics University of Crete, 71003 Heraklion, Greece.
\and
           Crete Center for Quantum Complexity and Nanotechnology, Department of Physics, University of Crete, 71003 Heraklion, Greece.
\and
APC, Univ Paris Diderot, Sorbonne Paris Cit\'e, APC, UMR 7164 CNRS, F-75205 Paris, France.
\and 
Department of Physics, McGill University\text{,} 3600 University Street\text{,} Montreal\text{,} QC\text{,} H3A 2T8\text{,} Canada.
\and
Theoretical Research Division, Nishina Center\text{,} RIKEN, Wako, Saitama 351-0198, Japan.
}

\abstract{%
   The thermal-photon emission from strongly coupled gauge theories at finite temperature is calculated by using holographic models for QCD in the Veneziano limit (V-QCD). These emission rates are then embedded in hydrodynamic simulations combined with prompt photons from hard scattering and the thermal photons from hadron gas to analyze the spectra and anisotropic flow of direct photons at RHIC and LHC. The results from different sources responsible for the thermal photons in the quark gluon plasma (QGP) including the weakly coupled QGP (wQGP) from perturbative calculations, strongly coupled $\mathcal{N}$=4 super Yang-Mills (SYM) plasma (as a benchmark for reference), and Gubser's phenomenological  model mimicking the strongly coupled QGP (sQGP) are then compared. It is found that the direct-photon spectra are enhanced in the strongly coupled scenario compared with the ones in the wQGP, especially at intermediate and high momenta, which improve the agreements with data. Moreover, by using IP-glassma initial states, both the elliptic flow and triangular flow of direct photons are amplified at high momenta ($p_T$>2.5 GeV) for V-QCD, while they are suppressed at low momenta compared to wQGP. The distinct results in holography stem from the blue-shift of emission rates in strong coupling. In addition, the spectra and flow in small collision systems were evaluated for future comparisons. It is found that thermal photons from the deconfined phase are substantial to reconcile the spectra and flow at high momenta.
}
\begin{flushright}
	CCQCN-2016-174\\
	CCTP-2016-16\\
	\vspace*{-0.8cm}
\end{flushright}
\maketitle
\section{Introduction}
\label{intro}
Thermal-photon production from the quark gluon plasma (QGP) phase plays an imperative role for direct photon production in heavy ion collisions. Particularly, recent measurements of large anisotropic flow of direct photons comparable to the hadron flow in RHIC \cite{Adare:2011zr} and LHC \cite{Lohner:2012ct} lead to a new puzzle. In general, electromagnetic probes barely interact with the QGP after they are produced and only record the local information in heavy ion collisions, although the scenario could be modified for long-lived plasmas such as the cosmic plasma \cite{Muller:2015maa,Yang:2015bva}. On the other hand, from most of theoretical predictions, the direct-photon spectra are underestimated as well. 
There have been intensive studies to reconcile the tension between experimental observations and theoretical predictions \cite{Shen:2013cca, Gale:2014dfa,PaquetShenDenicolEtAl2016,Kim:2016ylr}. Furthermore, some of novel mechanisms such as the enhanced thermal-photon production from strong magnetic fields were considered \cite{Basar:2012bp,Muller:2013ila}.  

 In most of theoretical approaches, the emission rate from the perturbative calculation with hard thermal loop resummation initiated by Arnold, Moore, and Yaffe (AMY) \cite{Arnold:2001ba,Arnold:2001ms} is applied for thermal photons generated from the QGP, where the next-to-leading order correction has been investigated in \cite{Ghiglieri:2013gia}. In order to facilitate our understandings for the impact from coupling dependence on the photons from QGP phase, one has to resort to non-perturbative approaches such as the AdS/CFT correspondence. An alternative approach by estimating the emission rate from the vector current correlator obtained by lattice simulations was recently proposed in \cite{Ghiglieri:2016tvj}.
 Although there have been considerable studies of thermal-photon production in holography such as  \cite{CaronHuot:2006te,Mamo:2013efa,Yee:2013qma,Wu:2013qja}, these studies do not incorporate the medium evolution and the photon production from other phases, which make the results difficult to be compared with experiments. In \cite{Iatrakis:2016ugz}, the authors thus 
 embed the photon emission rates from holography into medium evolution and include other contributions for direct-photon production. 

In \cite{Iatrakis:2016ugz}, two bottom-up holographic models are applied to model the sQGP, both of which break conformal invariance and match several  properties of QCD at finite temperature as calculated from lattice simulations. One of the models we employ was introduced by Gubser and Nellore (GN) \cite{Gubser:2008ny}. The other model is  V-QCD, which is a more sophisticated holographic model and it is based on the improved holographic QCD formalism \cite{Gursoy:2007cb,Gursoy:2007er}.
The flavor degrees of freedom are included by adding $N_f$ brane-antibrane pairs with a bulk tachyon field that describes the spontaneous breaking of chiral symmetry
\cite{Iatrakis:2010jb,Iatrakis:2010zf}. The photon-emission rates from these holographic models are then convoluted with the medium evolution. Furthermore, the contributions from prompt photons and thermal photons from hadron gas are incorporated to compute both the spectra and flow of direct photons in both RHIC and LHC energies.

In this proceeding, we review the major results found in \cite{Iatrakis:2016ugz}. The article is organized as following. In Sec. \ref{sec_2}, we present the photon-emission rates from holographic models  and discuss their difference in comparison with AMY. In Sec.\ref{sec_3}, we present numerical results of direct-photon spectra and anisotropic flow in comparison with experimental measurements, in which we further highlight the results for small collision systems. Finally, we make concluding remarks in Sec.\ref{sec_4}.    

\begin{figure}[h]
	\begin{center}
		{\includegraphics[width=7cm,height=5cm,clip]{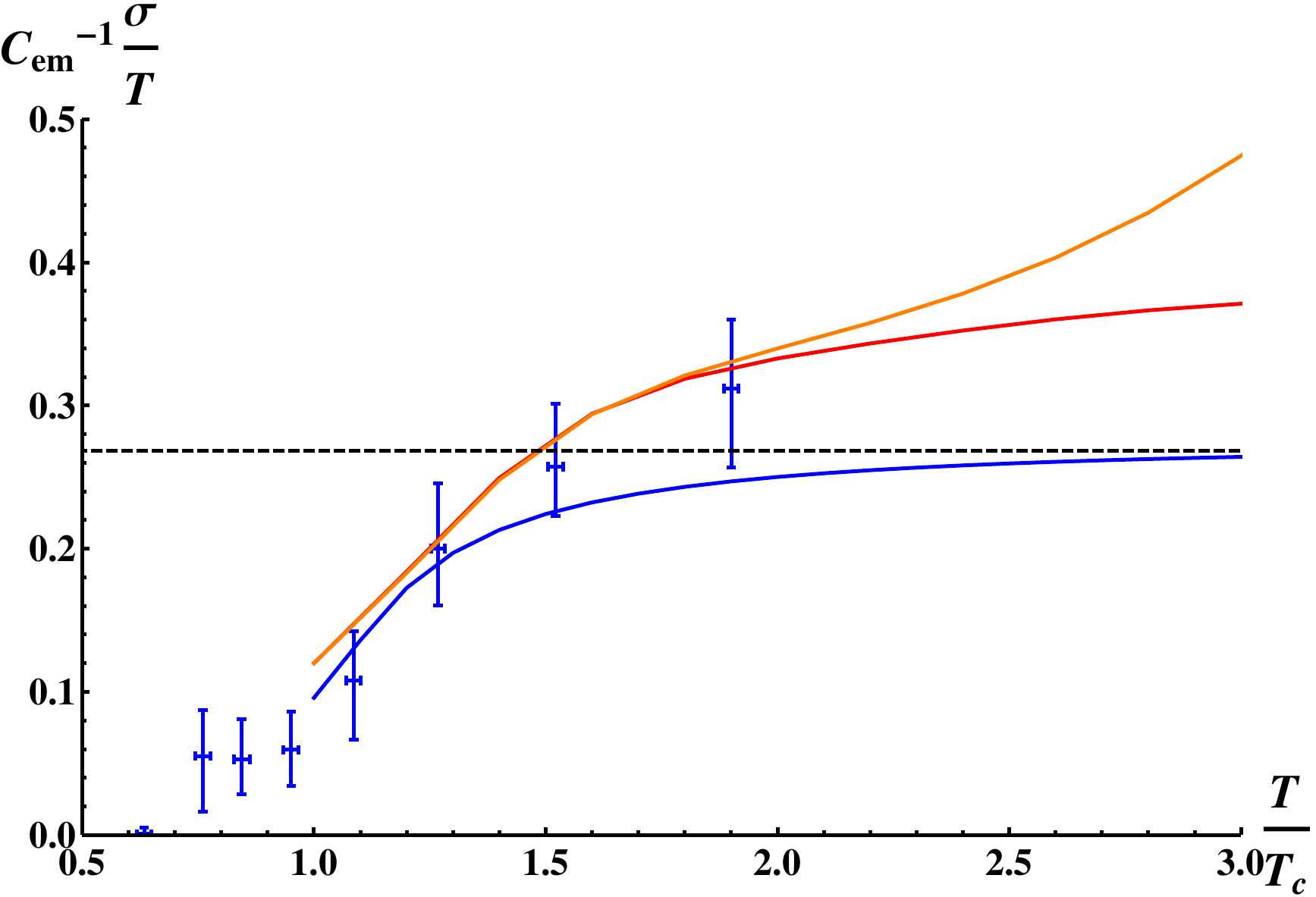}}
		\caption{Electric conductivity from different models \cite{Iatrakis:2016ugz}. The orange, red, and blue (from top to bottom ) curves correspond to VQCD1, VQCD2, and GN model, respectively. The black dashed line represents the result for the SYM plasma and the blue dots correspond to the lattice simulation with three flavors \cite{AartsAlltonAmatoEtAl2015}, where $C_{em}=2e^2/3$ and $N_c=N_f=3$.}\label{DC_conductivity}
	\end{center}
\end{figure}

\section{Electric Conductivity and Photon-Emission Rates}\label{sec_2}
In this section, we present the photon-emission rates from the GN model and V-QCD, while we will omit the technical details, which can be found in \cite{Iatrakis:2016ugz}. In general, the photon-emission rate is pertinent to the retarded correlator of currents in momentum space, 
\begin{eqnarray}
G_{\mu\nu}^{ab \,R}(k)=\int d^4(x-y)e^{ik\cdot(x-y)} \theta(x^0-y^0) \langle [ J^a_{\mu}(x), J^b_{\nu}(y) ] \rangle,
\end{eqnarray}
which can be evaluated by either holography or perturbative QCD. In thermal equilibrium, the photon production is then given by the light-like correlator,
\begin{eqnarray}
d\Gamma=-\frac{d^3k}{(2\pi)^3}\frac{e^2n_b(|{\bf k}|)}{|{\bf k}|}\text{Im}\left[\text{tr}\left(\eta^{\mu\nu}G_{\mu\nu}^{ab \, R}\right)\right]_{k^0=|{\bf k}|},
\end{eqnarray}
where $\Gamma$ denotes the number of photons emitted per unit time per unit volume and $n_b(|{\bf k}|)$ denotes the thermal distribution function for Bosons.
For describing photon emission in strongly coupled QCD, one has to construct suitable gauge-field-gravity couplings and choose proper normalization. Since the electric conductivity is governed by the infrared behavior of the current correlator, in \cite{Iatrakis:2016ugz}, such couplings and normalization are chosen to approximately fit the electric conductivity from lattice simulations. The results are shown in Fig.\ref{DC_conductivity}. For V-QCD, two types of couplings are considered, where one leads to saturation at high temperature as VQCD1 and one yields monotonic increase as VQCD2. 

\begin{figure*}[h]
	\centering
	\begin{tabular}{cc}
		\includegraphics[width=0.5\linewidth]{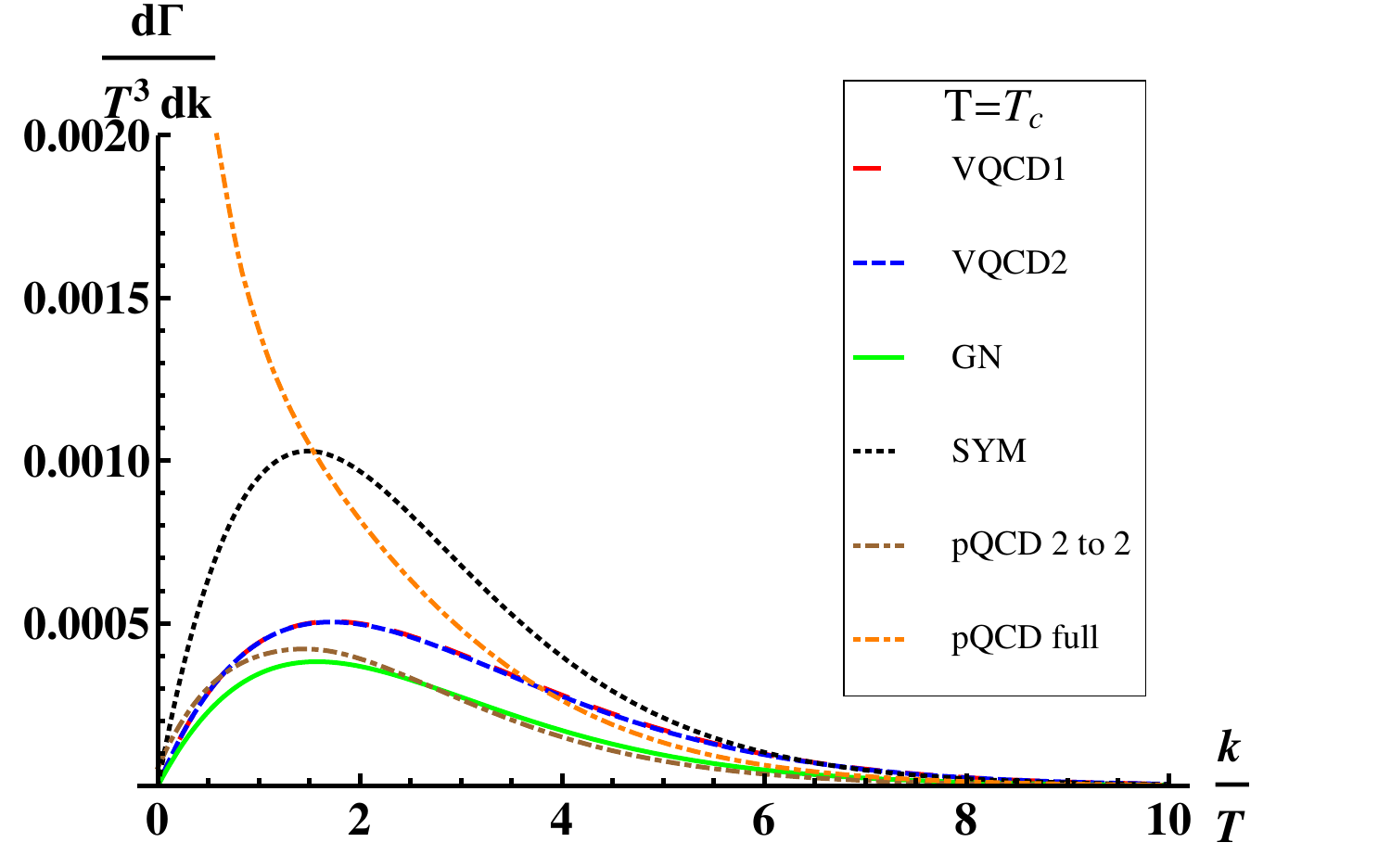} &
		\includegraphics[width=0.5\linewidth]{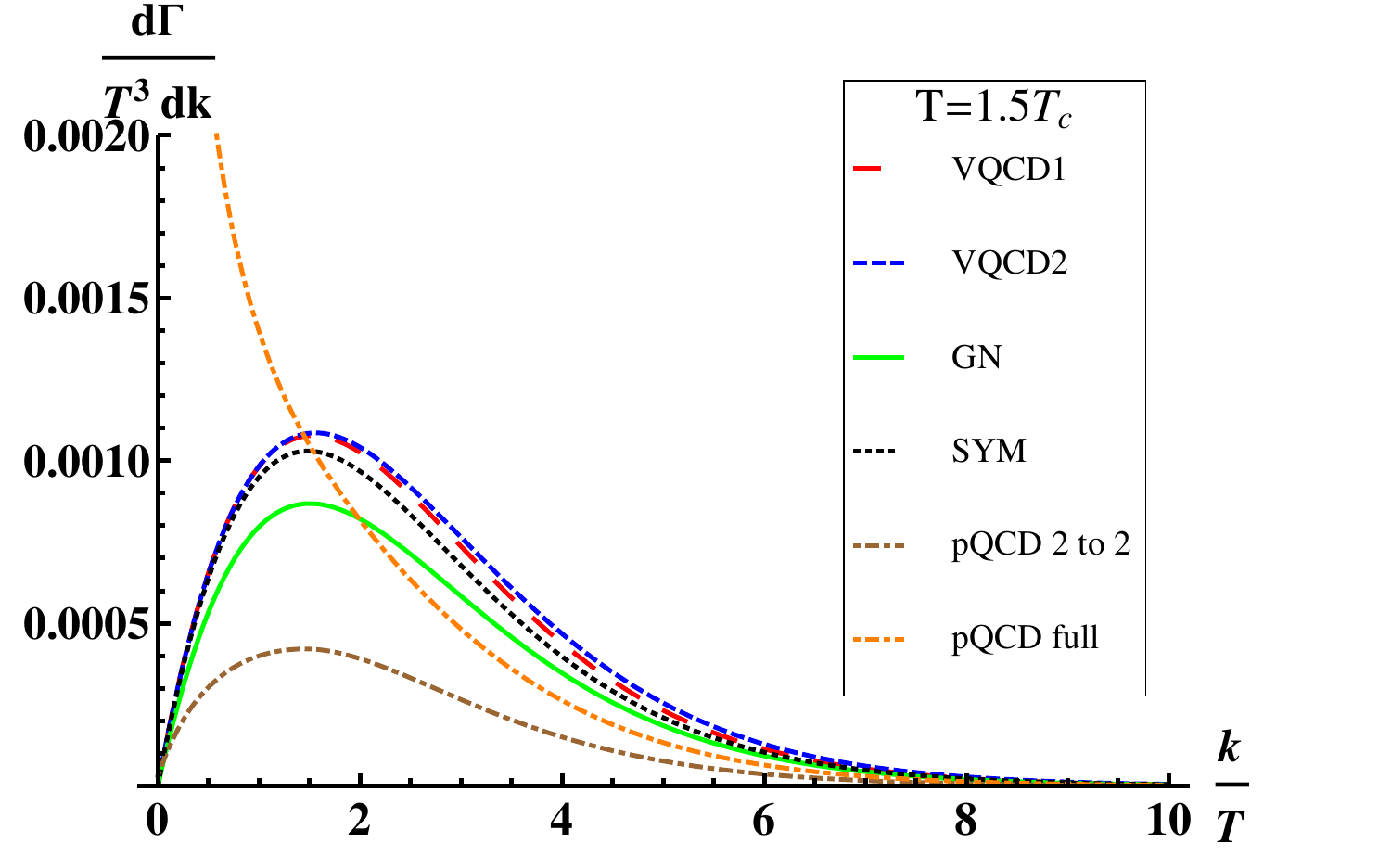} \\
		\includegraphics[width=0.5\linewidth]{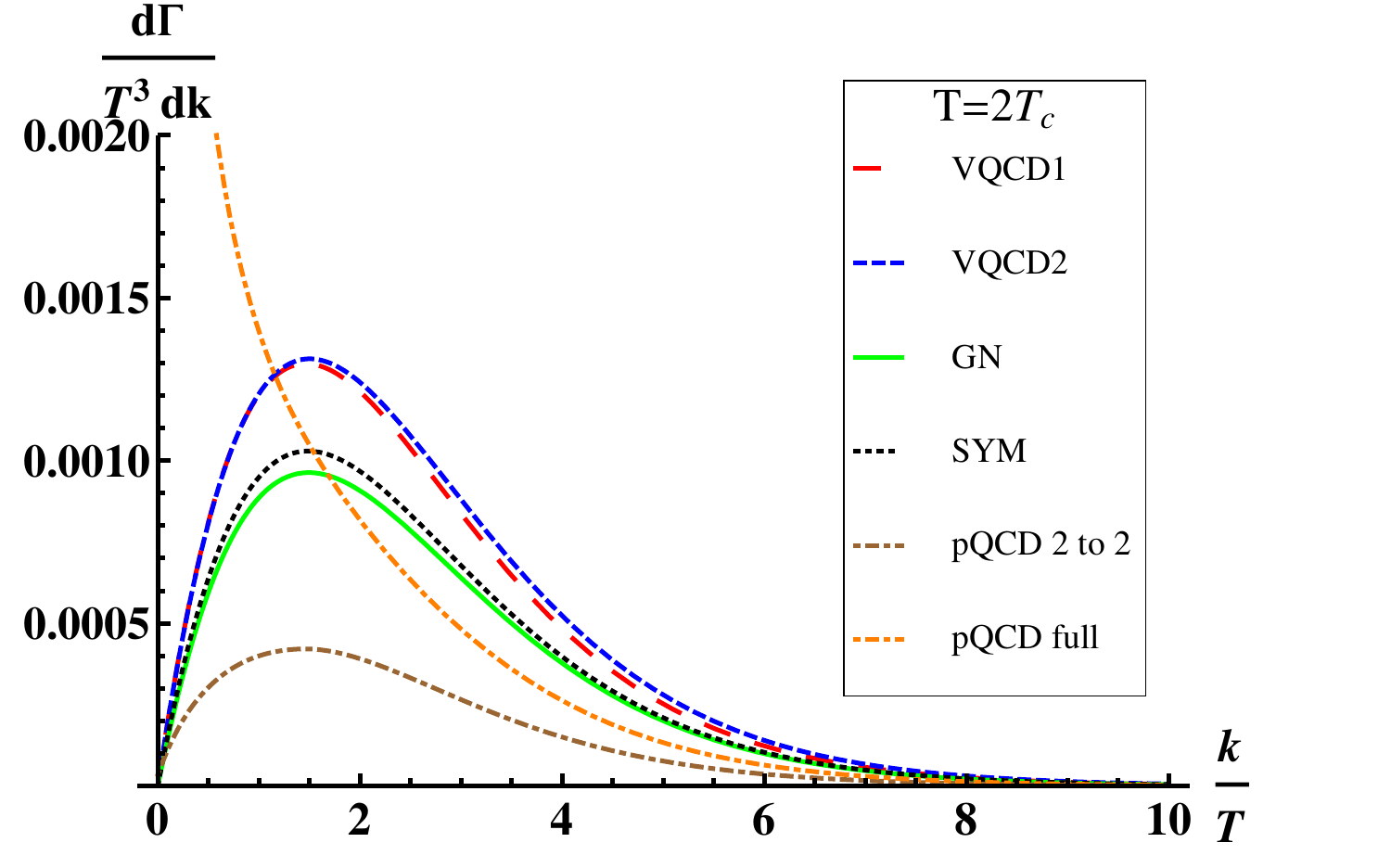} &
		\includegraphics[width=0.5\linewidth]{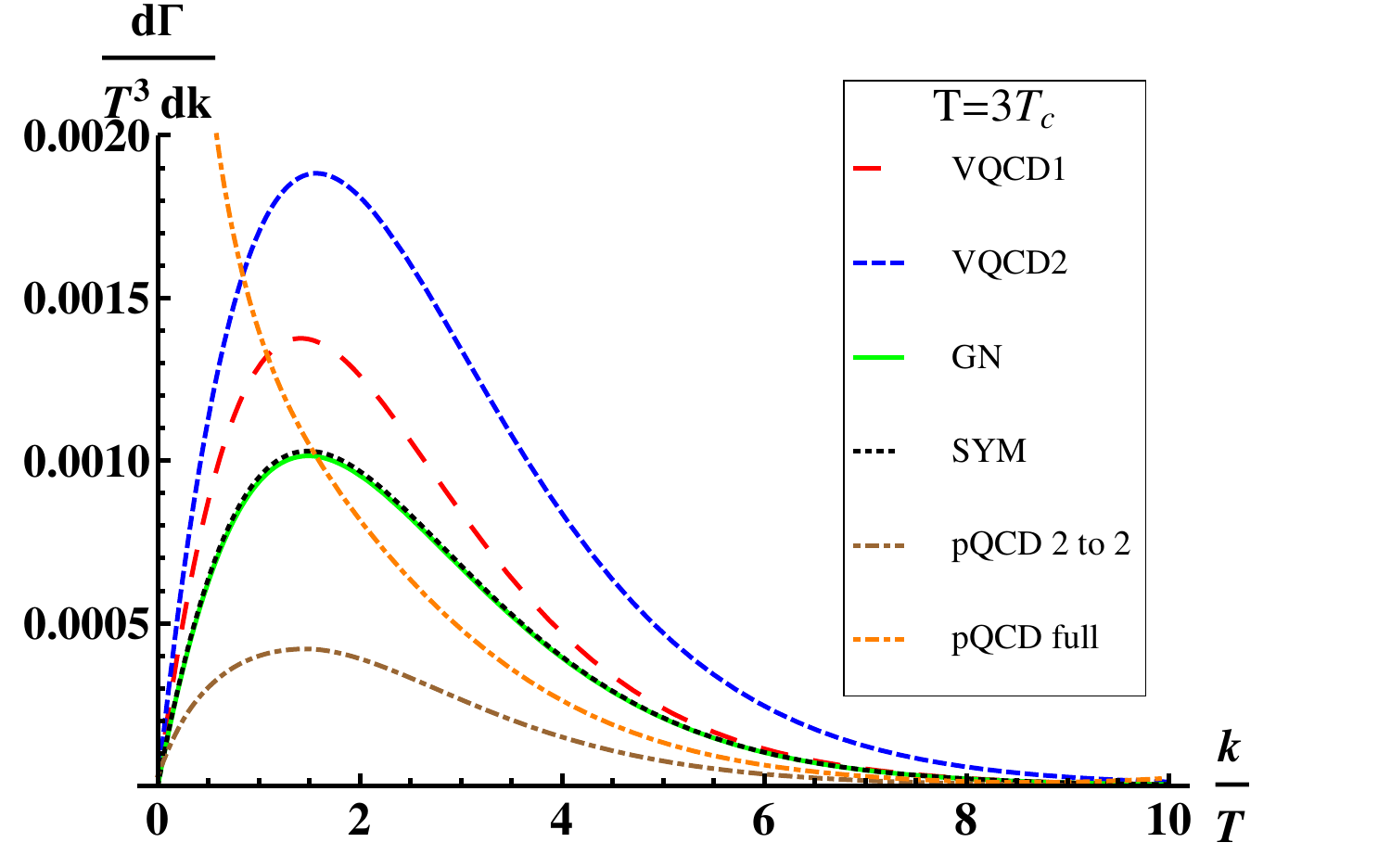}
	\end{tabular}
	\caption{Photon-emission rates from different models \cite{Iatrakis:2016ugz}. The pQCD rates are for the weakly coupled QGP at $g_s=2$. Here pQCD full correspond to 2 to 2 collisions plus collinear emissions\cite{Shen:2014nfa}.}
	\label{fig_photon_rates}
\end{figure*}

\begin{figure}[t]
	\begin{center}
		{\includegraphics[width=7cm,height=5cm,clip]{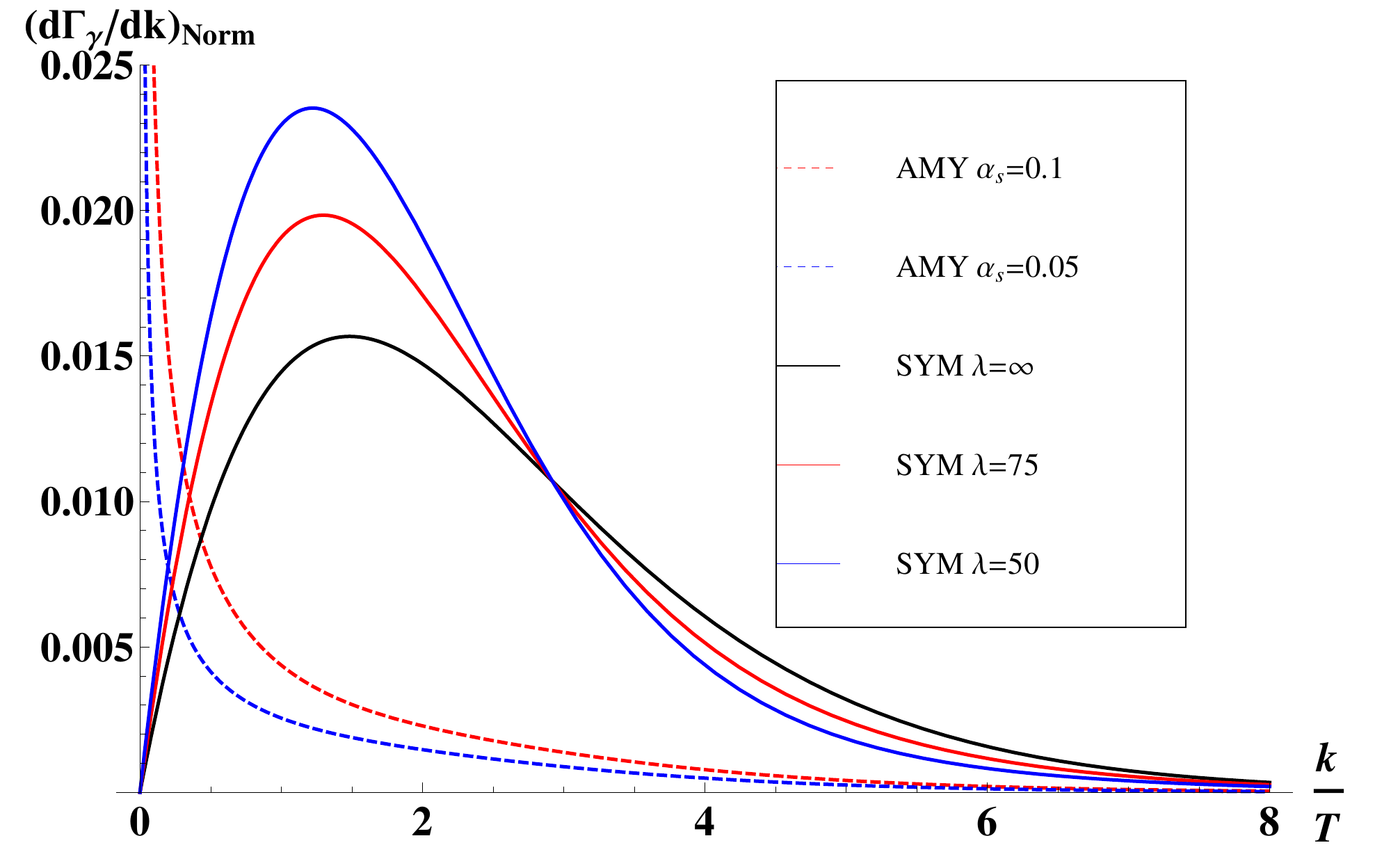}}
		\caption{The photon-emission rates normalized by $\alpha_{EM} N_c^2T^3$ for wQGP and strongly coupled SYM.}\label{blue_shift}
	\end{center}
\end{figure}

The corresponding photon-emission rates are shown in Fig.\ref{fig_photon_rates}. In general, similar to the finding in the SYM plasma \cite{CaronHuot:2006te}, the emission rates from holography models have distinct features at low energy compared with the ones in the weakly coupled scenario. However, it is worthwhile to note that the 2 to 2 collisions result in similar shapes of emission rates compared with those from holographic models, whereas the collinear emissions lead to IR divergence. In contrast, the photon production in holographic models incorporate both mechanisms, which "might" suggests that the collinear emissions are suppressed in strong coupling. As shown in the photon production from the SYM plasma with finite t'Hooft coupling $\lambda=g_s^2 N_c$ in holography \cite{Hassanain:2011ce}, the peak of the photon emission rate shifts to small $k/T$ when the coupling is reduced. This "blue-shift" of emission rates at strong coupling could also reduce the shear viscosity of photons in the QED+QCD plasma \cite{Yang:2015bva}. To highlight this "blue-shift", in Fig.\ref{blue_shift}, we further illustrate the comparison between the emission rates of strongly coupled $\mathcal{N}=4$ SYM with finite-coupling corrections \cite{Hassanain:2011ce} and the rates of pQCD from the phenomenological fits of AMY \cite{Arnold:2001ms}. Here the coupling dependence of the normalized rate in AMY is due to the logarithmic term related to the thermal mass, $\log(T/m_{\infty})\sim \log(1/g_s)$. Albeit the absence of smooth extrapolation between the results in strongly coupled and weakly coupled regions, it is tentative to conjecture that such "blue-shift" is universal, which entails future efforts in higher-order corrections from both perturbative and non-perturbative regions.           

\section{Direct-Photon Spectra and Flow}\label{sec_3} 
The holographic models discussed above are only responsible for thermal photons in the QGP phase. To make direct comparisons with experimental data, one has to be embedded them into the medium evolution and include both prompt and thermal photons. The prompt photons come from hard interactions in early times and the thermal photons incorporate the contributions from the QGP phase and hadronic phase. The momentum distribution of thermal photons is computed by first producing photons in the local rest of frame of every fluid cell,
whose temperature $T(x)$ is higher than the system's freeze-out temperature $T_\mathrm{freezeout}$. Then these photons are boosted with the corresponding fluid velocity $u(x)$
to the lab frame,
\begin{equation}
q \frac{dN_\mathrm{thermal}^\gamma}{d^3 q} = \int_{T > T_\mathrm{freezeout}} d^4 x \bigg[ q \frac{dR^\gamma}{d^3 q}\left(T(x), E_q \right)\bigg\vert_{E_q = q \cdot u(x)} \bigg],
\end{equation}
where the thermal photon emission rate is denoted as $q \frac{dR^\gamma}{d^3 q}$. The direct photon spectra is the sum of thermal and prompt photons,
\begin{equation}
q \frac{dN_\mathrm{direct}^\gamma}{d^3 q} = q \frac{dN_\mathrm{thermal}^\gamma}{d^3 q} + q \frac{dN_\mathrm{prompt}^\gamma}{d^3 q}.
\end{equation}
More details of the model setup are discussed in \cite{Iatrakis:2016ugz}. In the following subsections, we highlight the important findings therein. 

\begin{figure*}[h!]
	\centering
	\begin{tabular}{cc}
		\includegraphics[width=0.48\linewidth,height=0.58\linewidth]{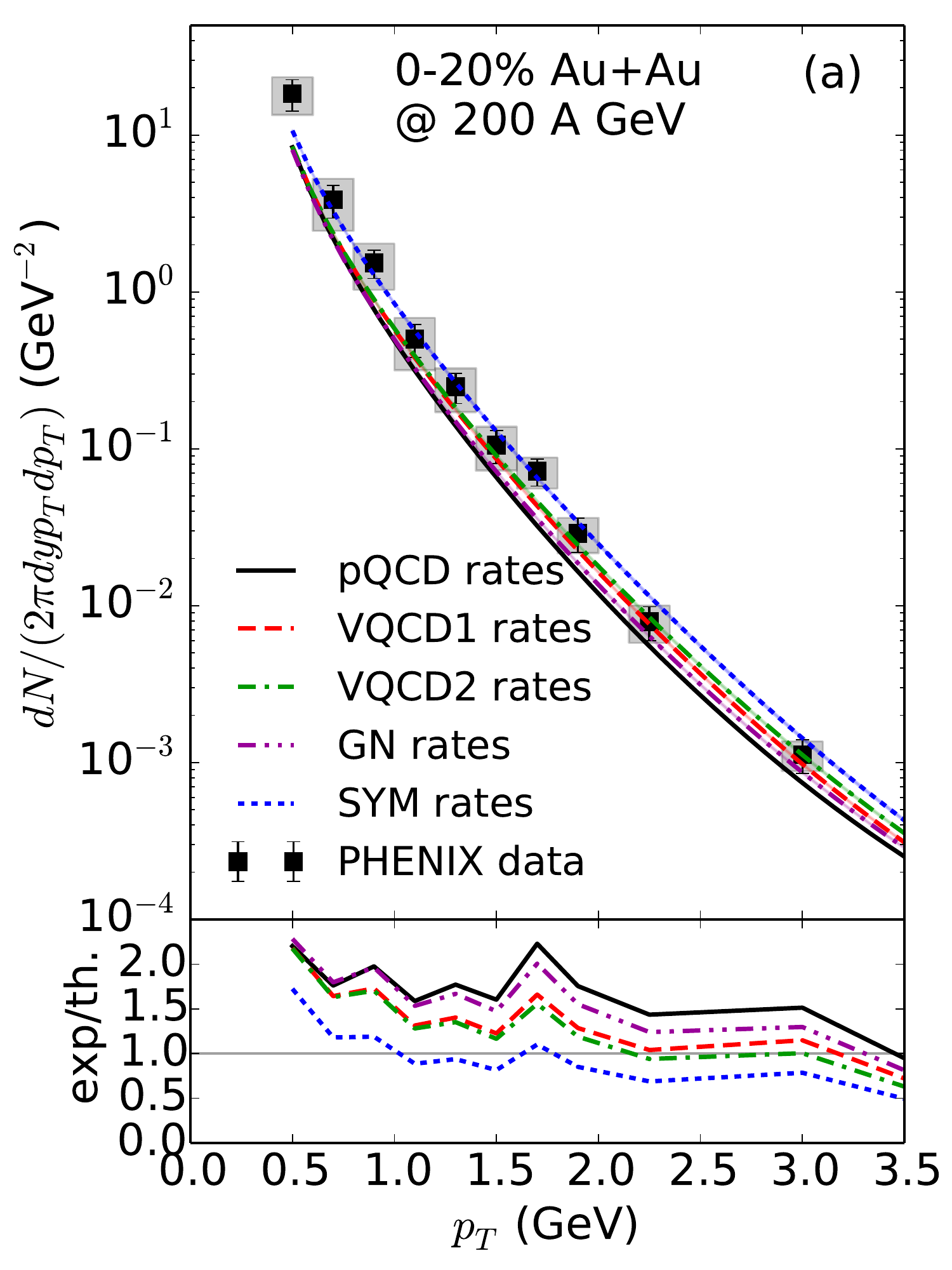} &
		\includegraphics[width=0.48\linewidth,height=0.58\linewidth]{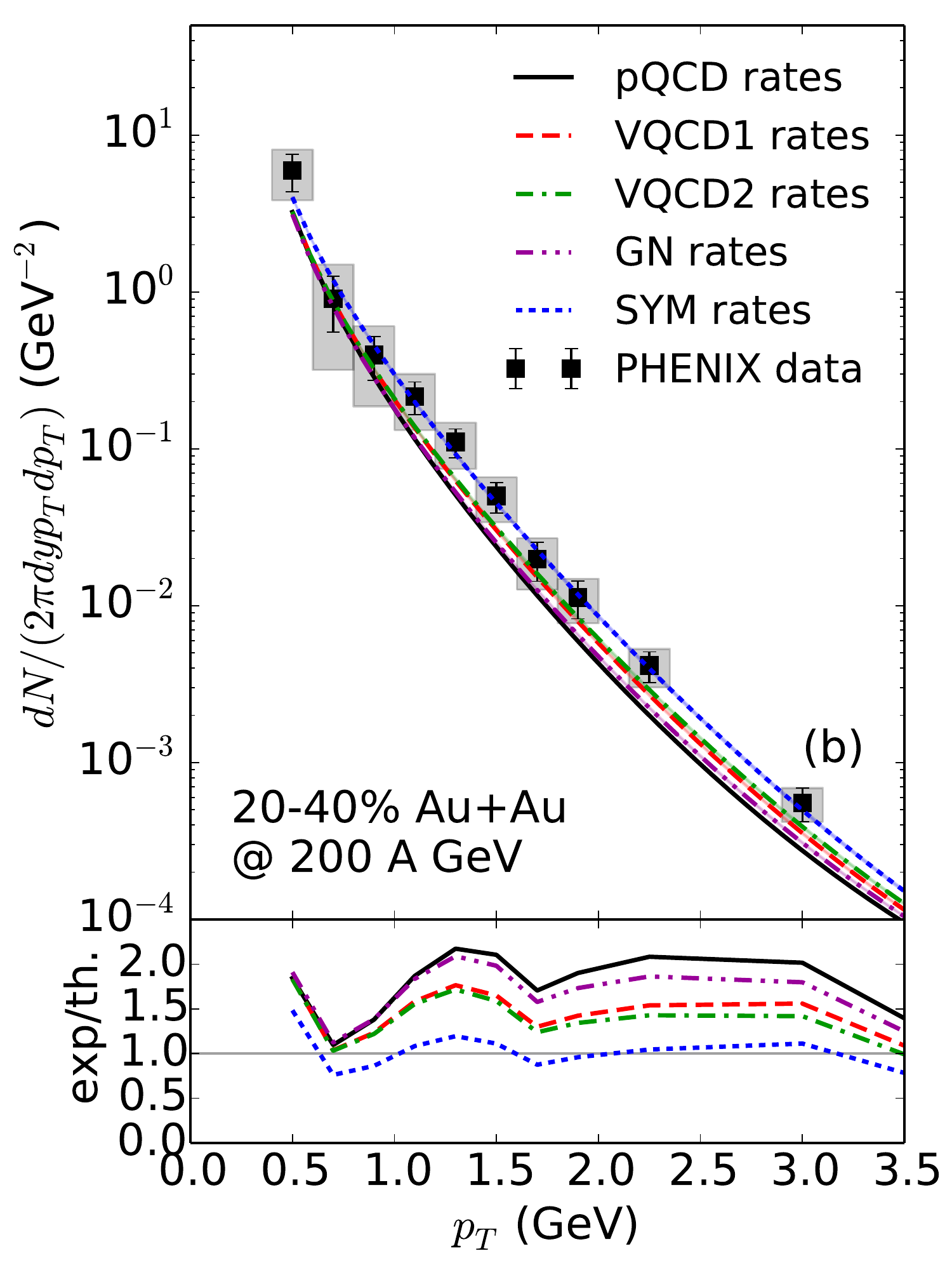} \\
		\includegraphics[width=0.48\linewidth,height=0.58\linewidth]{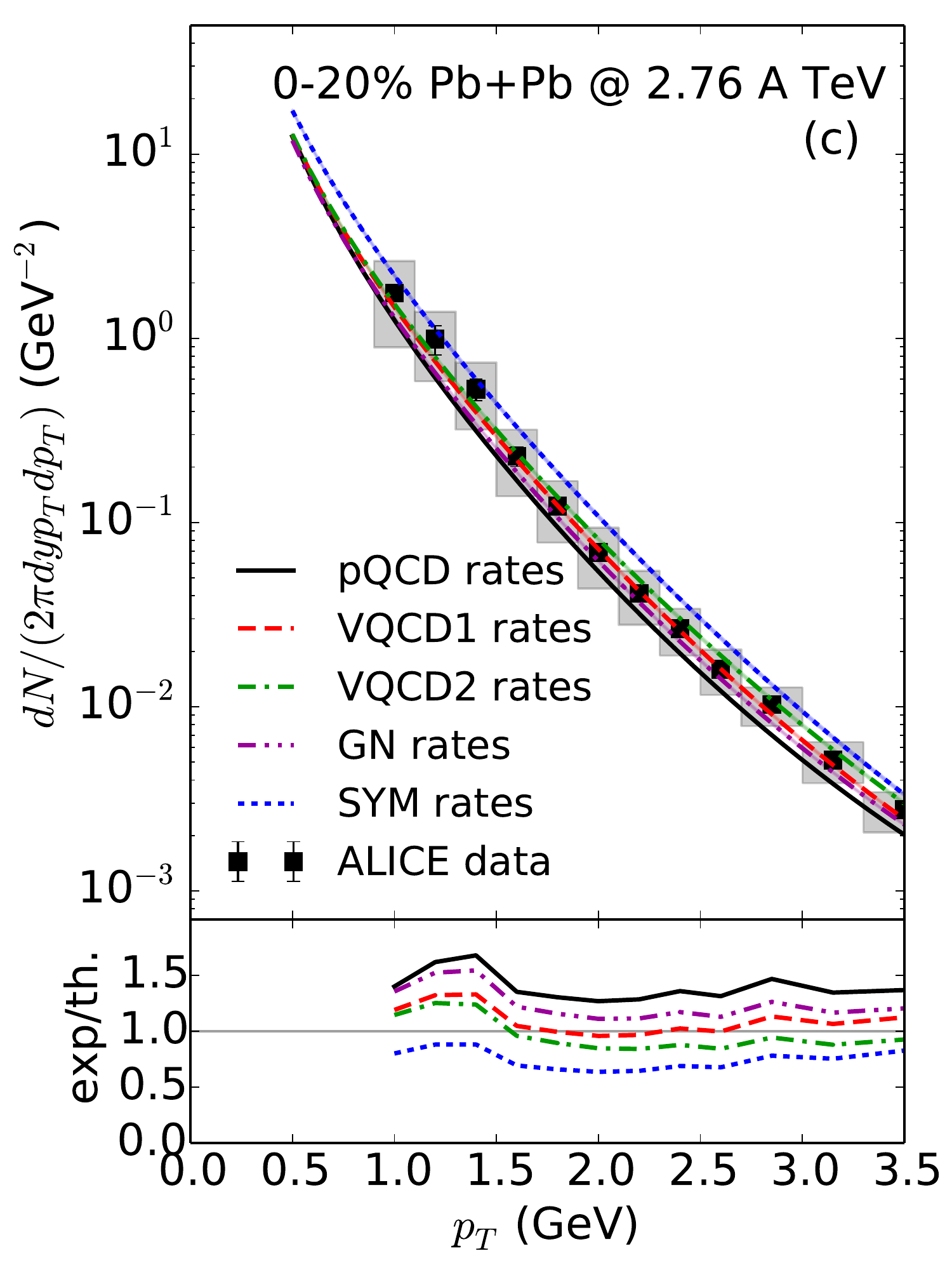} &
		\includegraphics[width=0.48\linewidth,height=0.58\linewidth]{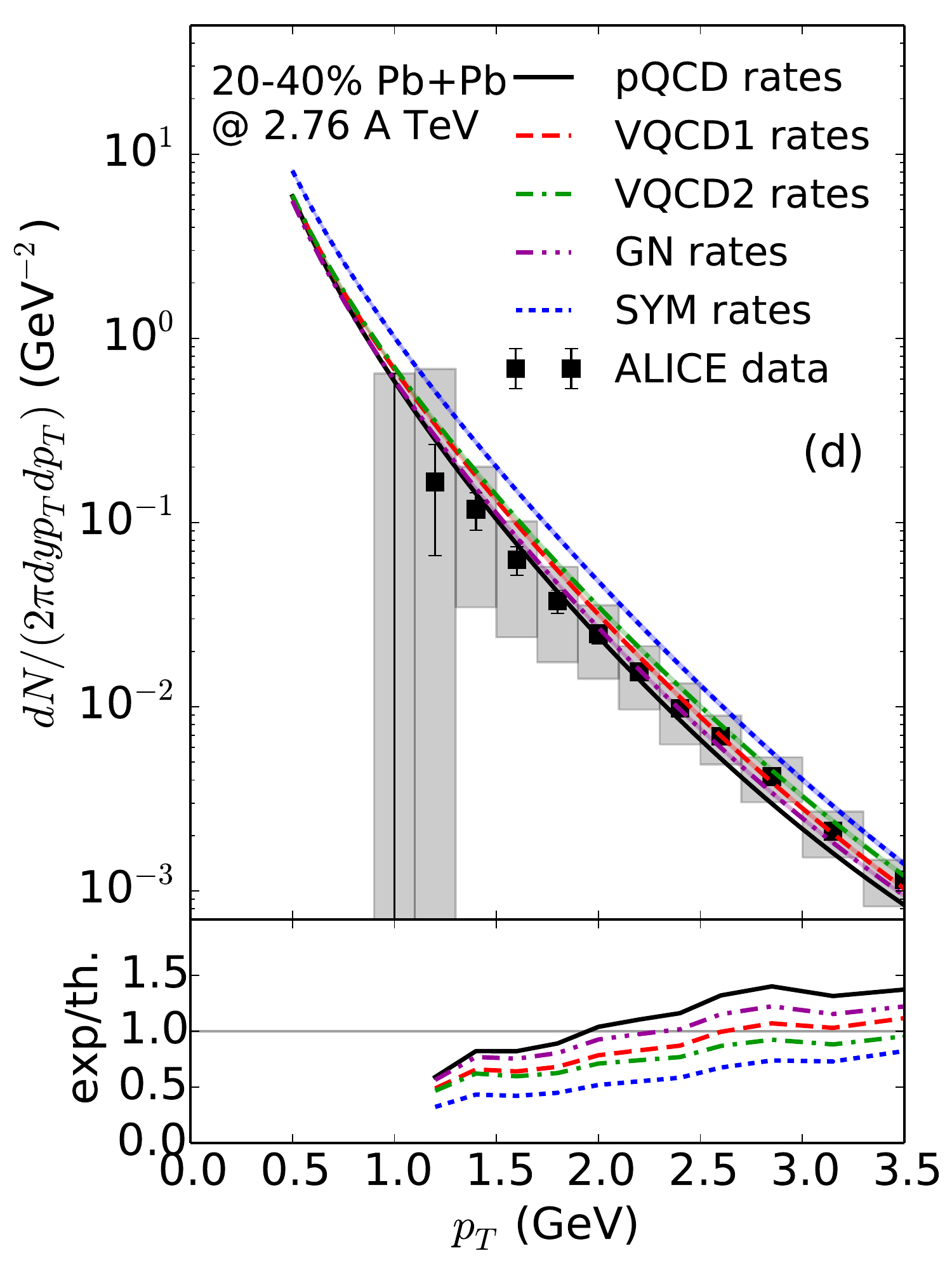}
	\end{tabular}
	\caption{Results for direct-photon spectra \cite{Iatrakis:2016ugz}. (Color online) Direct photon spectra from 0-20\% (a) and 20-40\% (b) Au+Au collisions at 200 GeV compared with the PHENIX measurements \cite{Adare:2014fwh} and from 0-20\% (c) and 20-40\% (d) Pb+Pb collisions at 2.76 $A$\,TeV compared with the ALICE measurements \cite{Adam:2015lda}. The ratios of experimental data to theoretical results are shown in the bottom of each plot.}
	\label{fig6.1}
\end{figure*}

%
\begin{figure*}[h!]
	\centering
	\begin{tabular}{cc}
		\includegraphics[width=0.48\linewidth]{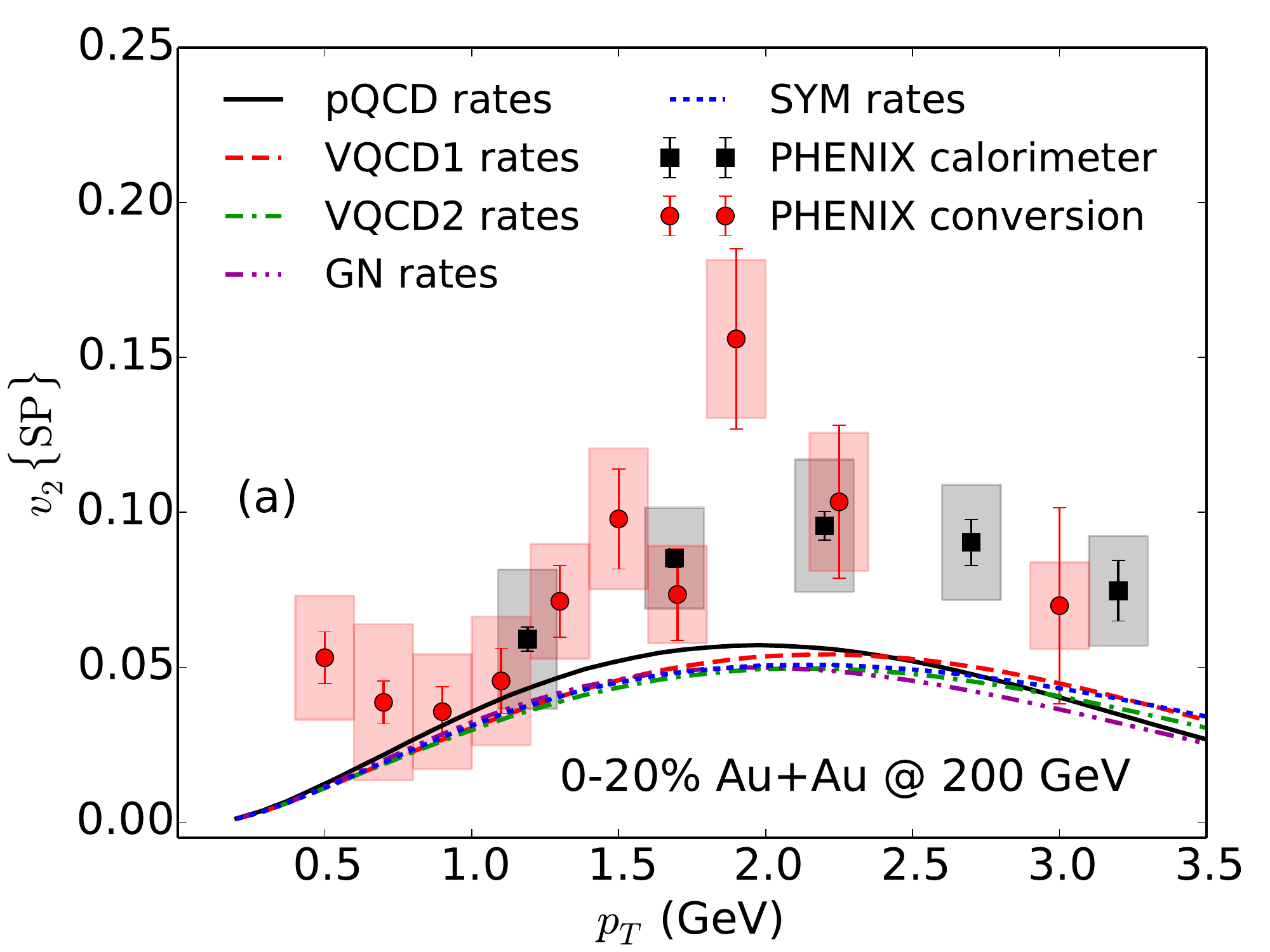} &
		\includegraphics[width=0.48\linewidth]{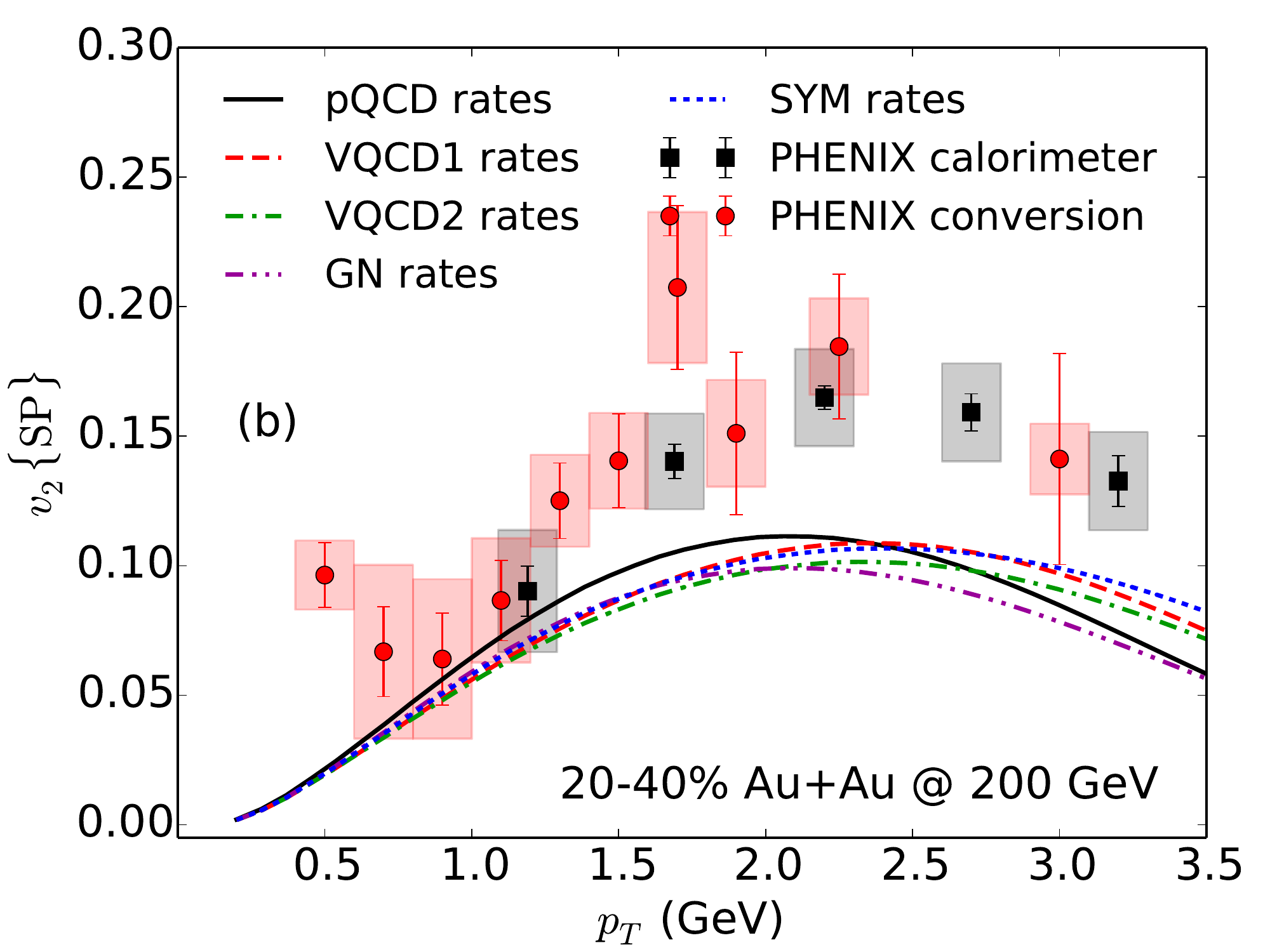} \\
		\includegraphics[width=0.48\linewidth]{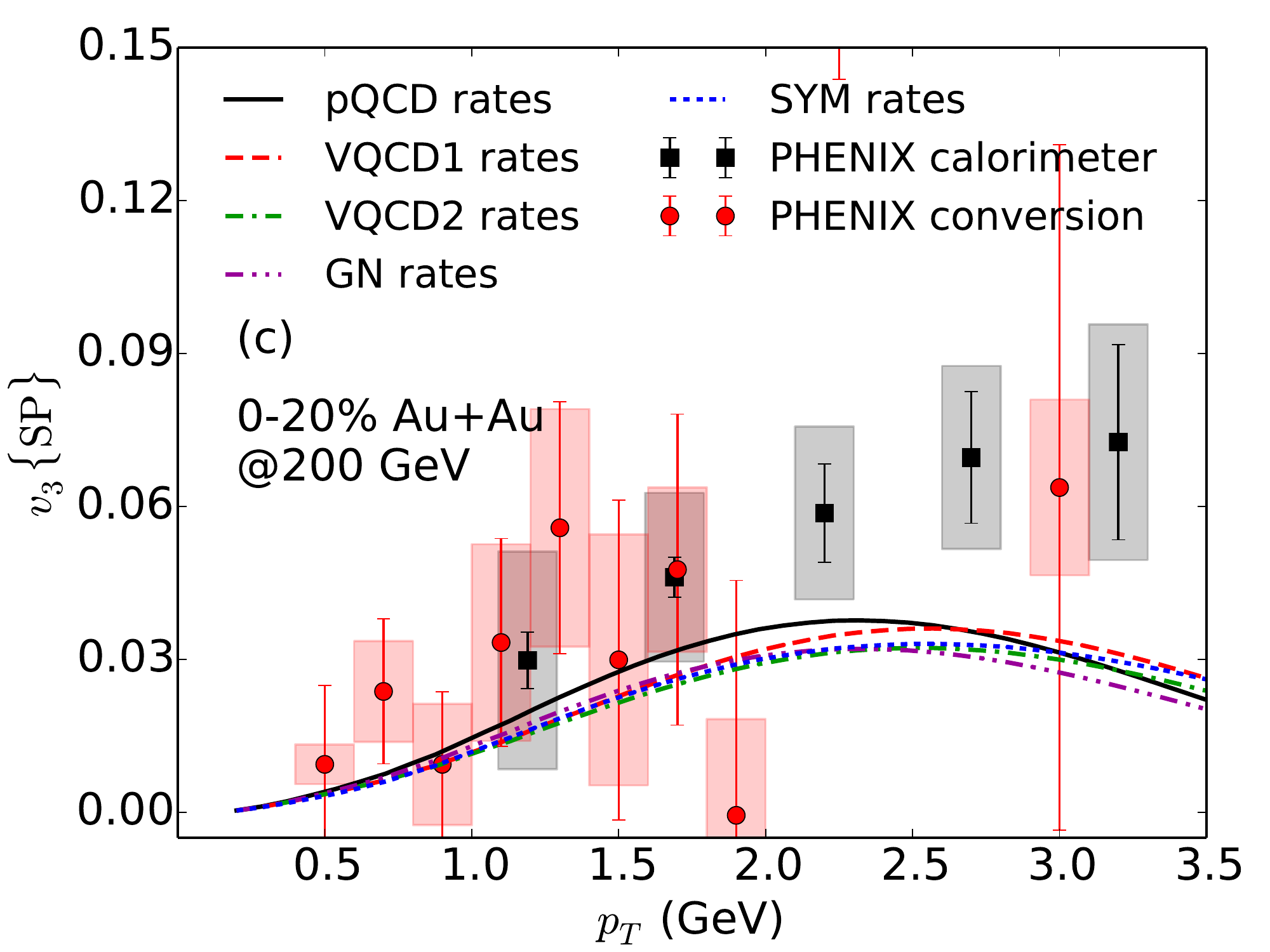} &
		\includegraphics[width=0.48\linewidth]{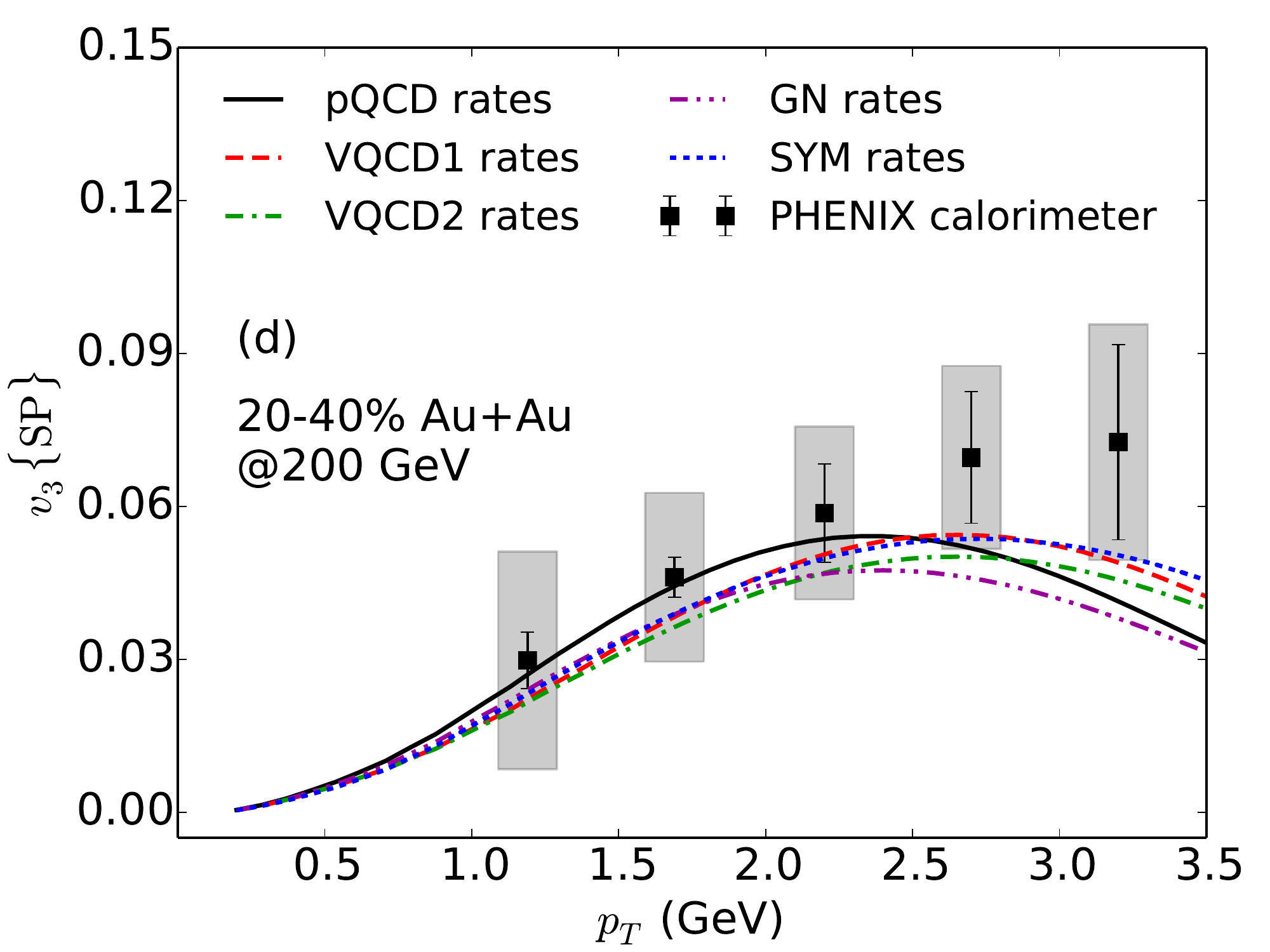}
	\end{tabular}
	\caption{Results for anisotropic flow in RHIC \cite{Iatrakis:2016ugz}. (Color online) Direct photon anisotropic flow $v_{2,3}$ from 0-20\% (a,c) and 20-40\% (b,d) Au+Au collisions at 200 GeV compared with the PHENIX measurements \cite{Adare:2015lcd}.  }
	\label{fig6.2}
\end{figure*}
\begin{figure}[h!]
	\centering
	\includegraphics[width=0.5\linewidth]{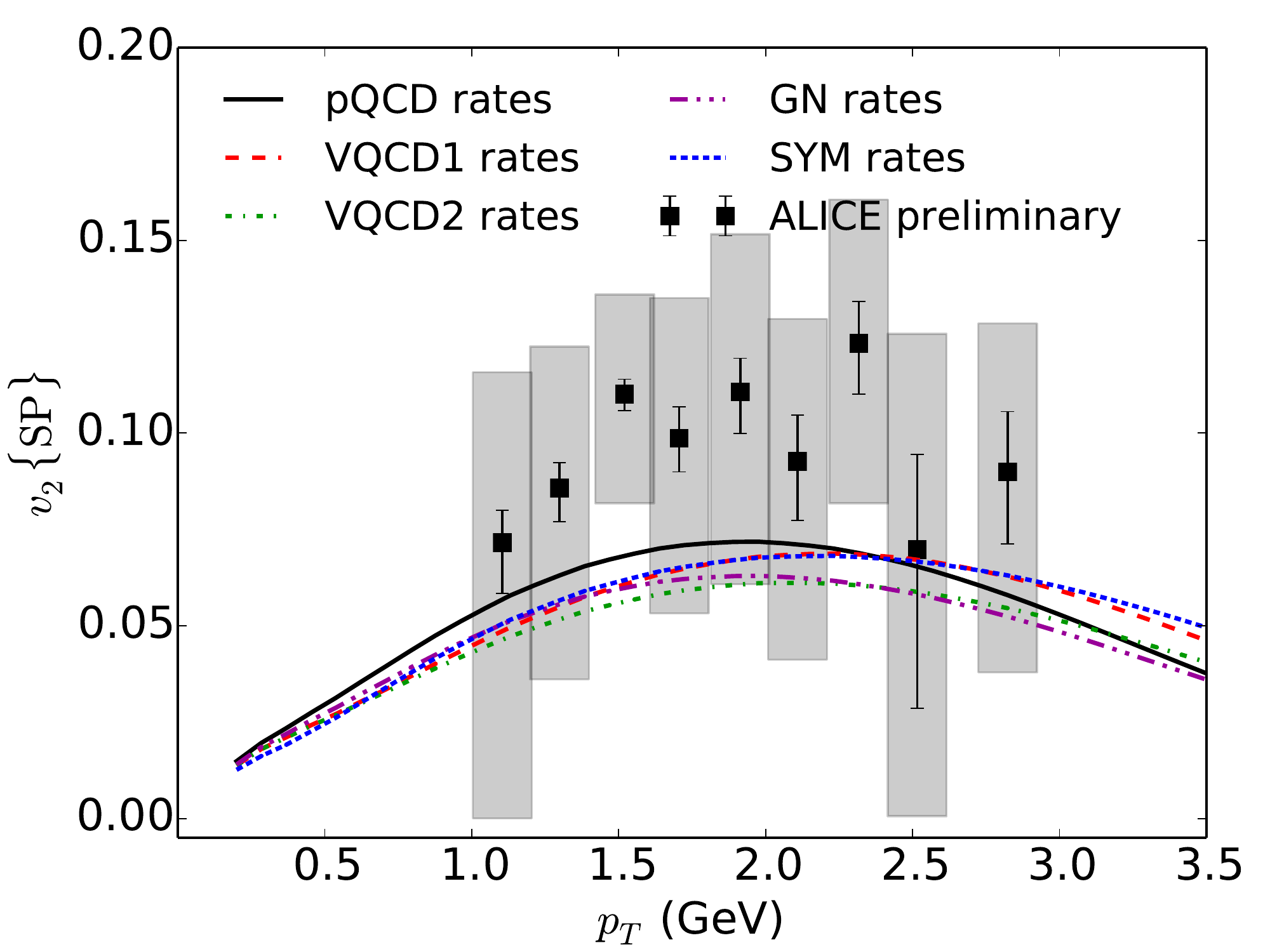}
	\caption{The result for anisotropic flow in LHC \cite{Iatrakis:2016ugz}. (Color online) Direct photon anisotropic flow $v_{2}$ in 0-40\% Pb+Pb collisions at 2.76 $A$\,TeV compared with the ALICE measurements \cite{Lohner:2012ct,Lohner:2013thesis}. }
	\label{fig6.3}
\end{figure}

\subsection{Direct photon spectra in relativistic heavy-ion collisions}

Direct photon spectra using different sets of QGP photon rates are compared to the experimental measurements in Au+Au collisions at 200 $A$\,GeV at the RHIC and in Pb+Pb collisions at 2.76 $A$\,TeV at the LHC in Figs.~\ref{fig6.1}.
The QGP photon rates from holographic models result in more thermal radiation compared to the results with the pQCD rate. 
The reasons for this depend on the holographic model.
First, at strong coupling we expect more photon emissions than at weak coupling. On top of this, SYM which contains extra supersymmetric partners is expected to give the highest rate and this is turns out to be correct. The GN and VQCD models are non supersymmetric and have the same number of (perturbative) degrees of freedom as QCD. 

Therefore, among the different holographic rates, the SYM rates give the most direct photons.
Although the electric conductivities in the two VQCD models exceed the one in SYM model for $T > 1.5 T_c$, the direct photon yields from the VQCD models are smaller compared to the SYM results. This is because that most of the thermal photon radiations are coming from the phase transition region, $150 < T < 250$ MeV, where the space-time volume is the largest. In this temperature region, the photon emission rates are suppressed in the VQCD models compared to the SYM rates. On the other hand, the GN model leads to smaller spectra compared to the ones for VQCD and SYM models as expected from the electric conductivity and emission rates.

\subsection{Direct photon anisotropic flow coefficients }
On the one hand, the absolute yield of direct photon spectra provides  information about the system's space-time volume as well as the degrees of freedom of photon emitters in the medium. On the other hand, the anisotropic flows of direct photons are more sensitive to the relative temperature dependence of photon rates and their interplay with the development of hydrodynamic anisotropic flows during the evolution.

In Figs.~\ref{fig6.2} and \ref{fig6.3}, direct photon anisotropic flow coefficients, $v_{2,3}\{\mathrm{SP}\}(p_T)$(with scalar-product method), are shown at the RHIC and LHC energies together with the experimental measurements. Since the underlying hydrodynamic medium is kept fixed for all sets of calculations, we here show curves without statistical error bands for better visual comparisons. At both collision energies, the weakly-coupled QCD rates gives the largest direct photon $v_n$ in the intermediate $p_T$ region, $1 < p_T < 2.5$ GeV. At the higher $p_T > 3.0$ GeV, the $v_n$ results using holographic rates are larger. This interesting hierarchy of direct photon $v_{2,3}$ is a results of the interplay between the temperature dependence of emission rates and the space-time structure of the hydrodynamic flow distribution.

\subsection{Direct photon emission in small collision systems}

Recently, sizable thermal radiation was found in high multiplicity light-heavy collisions in Ref.~\cite{Shen:2015qba}. Owing to large pressure gradients, small collision systems, such as p+Pb and d+Au collisions, expand more rapidly compared to the larger Au+Au collisions, which yield higher freeze-out temperature. This leads to a smaller hadronic phase in these collision systems. Most of the thermal photons come from the hot QGP phase, $T > 180$ MeV \cite{Shen:2015qba}. Hence, the difference between the QGP photon emission rates should be more distinctive in these small collision systems.

\begin{figure}[h!]
	\centering
	\begin{tabular}{cc}
		\includegraphics[width=0.48\linewidth]{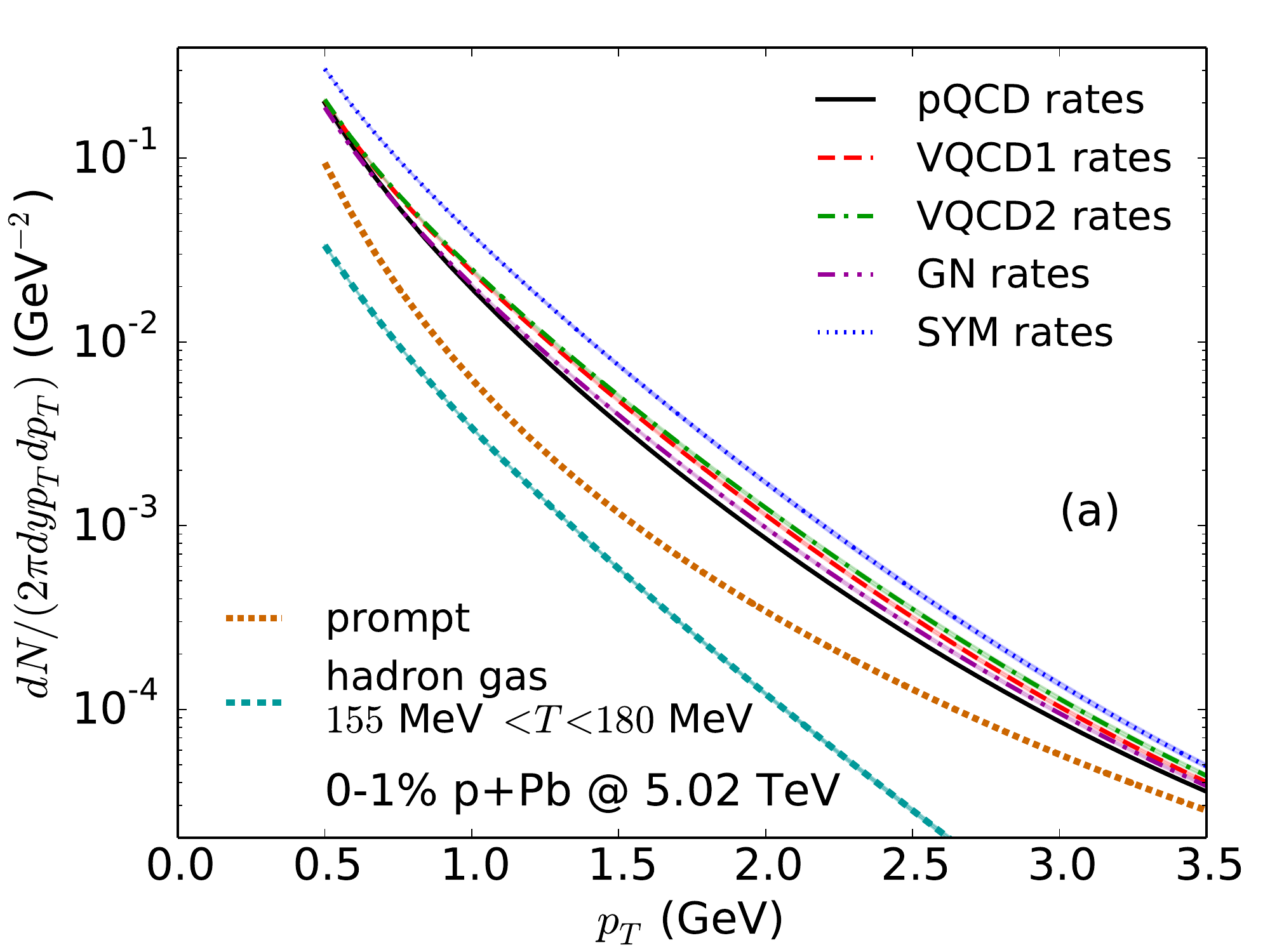} &
		\includegraphics[width=0.48\linewidth]{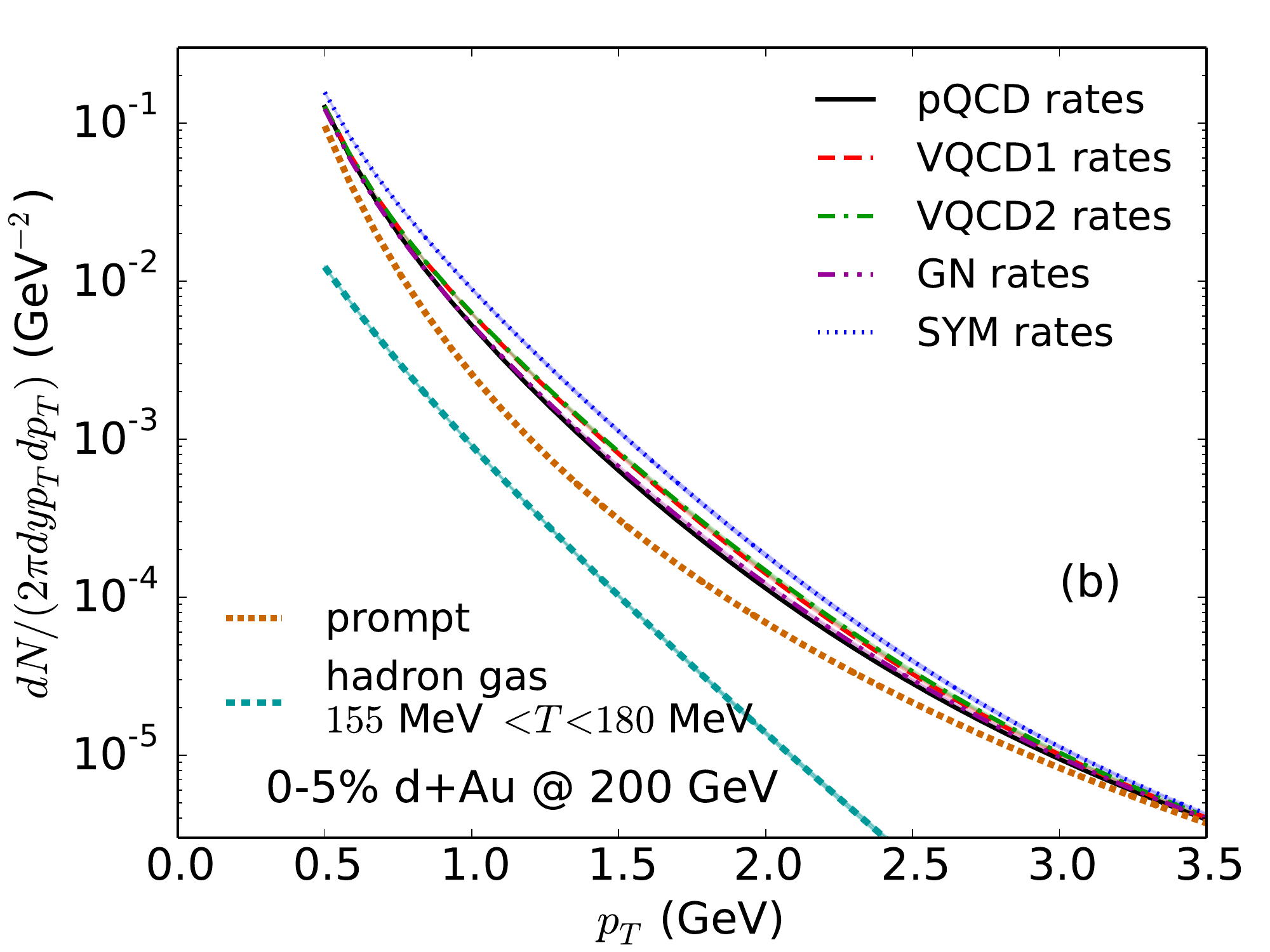} \\
		\includegraphics[width=0.48\linewidth]{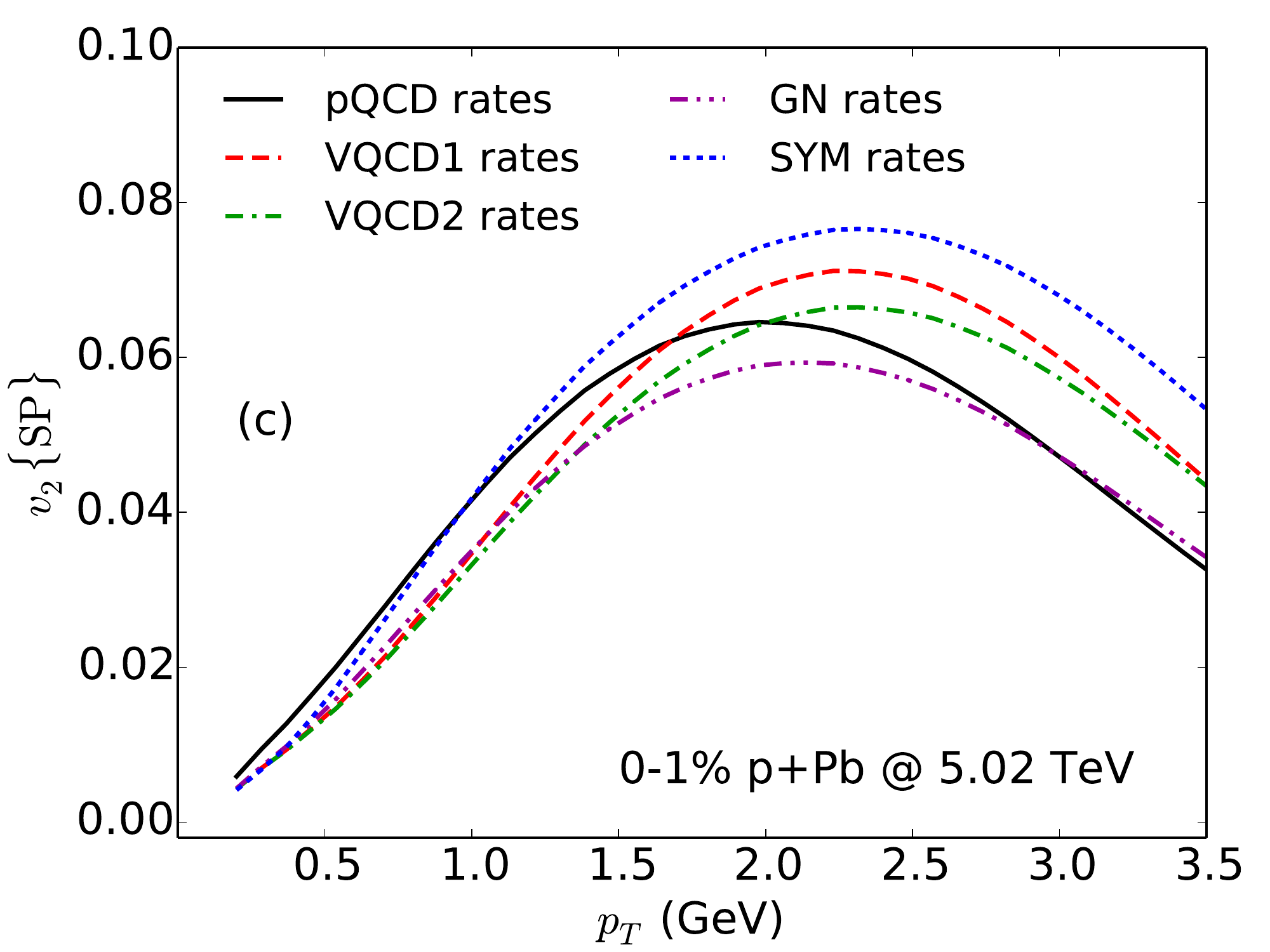} &
		\includegraphics[width=0.48\linewidth]{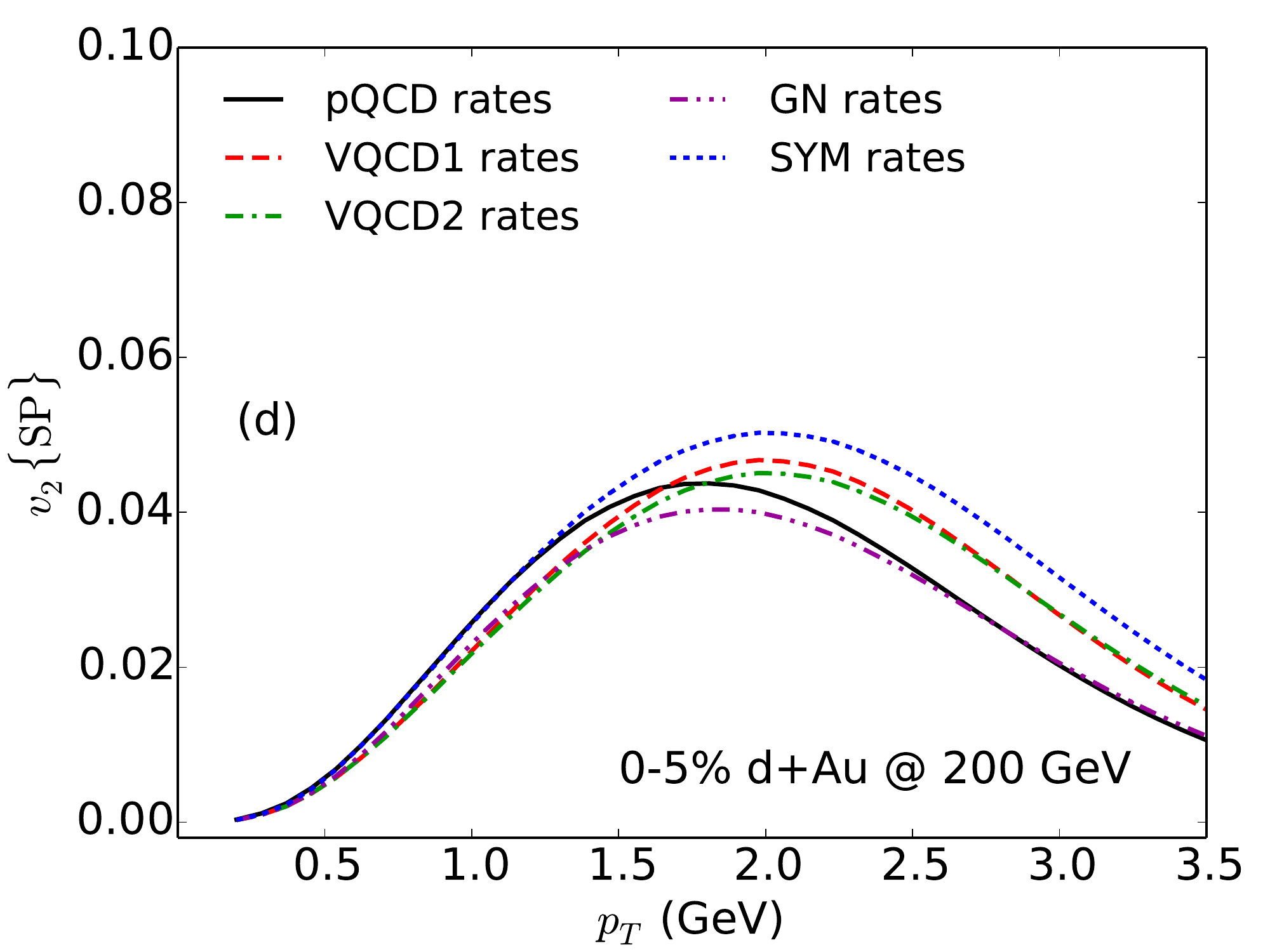} \\
		\includegraphics[width=0.48\linewidth]{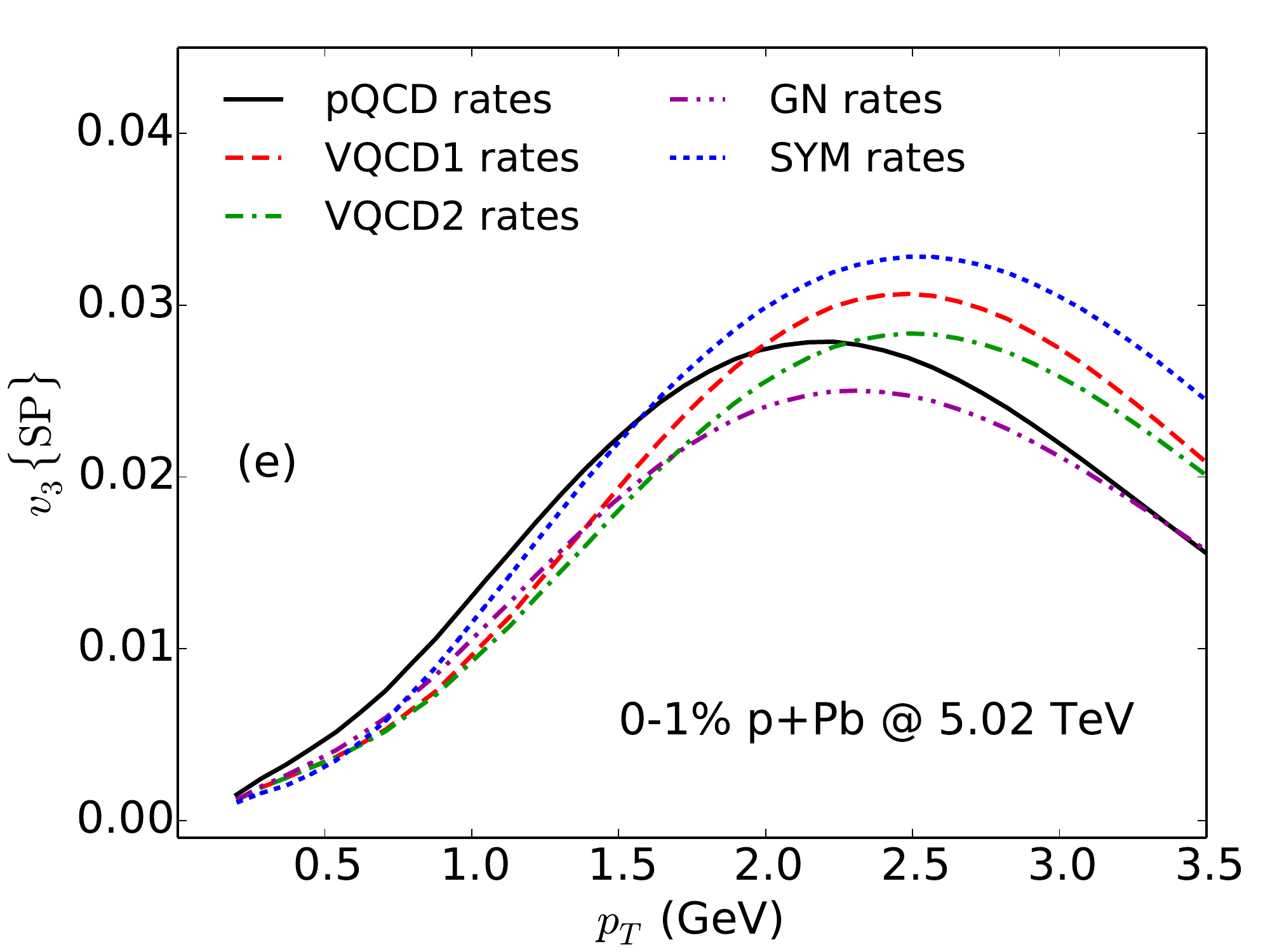} &
		\includegraphics[width=0.48\linewidth]{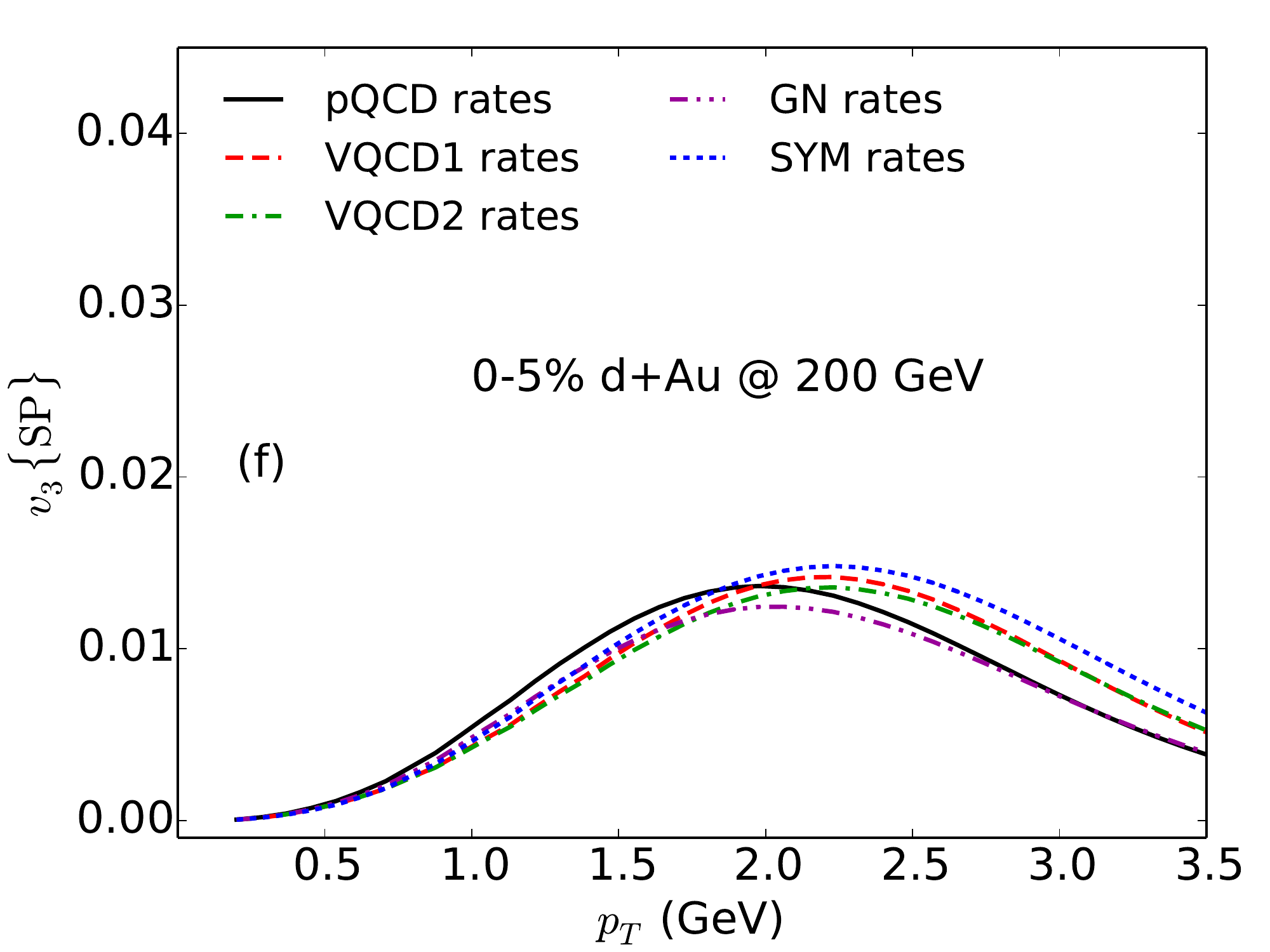}
	\end{tabular}
	\caption{Results for small collision systems \cite{Iatrakis:2016ugz}. (Color online) Direct photon spectra from 0-1\% p+Pb collisions at 5.02 TeV $(a)$ and 0-5\% d+Au collisions at 200 GeV $(b)$ using different sets of emission rate in the QGP phase. Direct photon anisotropic flow coefficients, $v_{2,3}(p_T)$, in 0-1\% p+Pb collisions at 5.02 TeV $(c,e)$ and 0-5\% d+Au collisions at 200 GeV $(d,f)$ using different sets of emission rate in the QGP phase.}
	\label{fig6.4}
\end{figure}
In contrast to nucleus-nucleus collisions that were analyzed  in the previous sections, full (3+1)D hydrodynamic simulations with Monte-Carlo Glauber model as initial conditions are employed for the medium evolution in central p+Pb and d+Au collisions \cite{Shen:2016zpp}. The parameters in the hydrodynamic model are chosen such that a variety of hadronic flow observables can be reproduced. Starting at an initial proper time $\tau_0 = 0.6$ fm/$c$, every fluctuating energy density profile is evolved in full 3+1 dimensions with $\eta/s = 0.08$ for d+Au collisions or with $\eta/s = 0.10$ for p+Pb collisions.

In \cite{Iatrakis:2016ugz}, this calibrated hydrodynamic medium is applied to study the sensitivity of direct photon observables in small systems to the different sets of QGP photon emission rates.

In Figs.~\ref{fig6.4}, direct photon spectra and their anisotropic coefficients are shown for top 0-1\% p+Pb collisions at 5.02 TeV and 0-5\% d+Au collisions at 200 GeV. In high-multiplicity events of small collision systems, thermal radiation can reach up to  a factor of 2 of the prompt contribution. Similar to nucleus-nucleus collisions, the direct photon spectra using the emission rates that are derived from strongly coupled theory are larger than the QCD rates. 

The difference is smaller in the d+Au collisions compared to p+Pb collisions at the higher collision energy. Although the hierarchy of direct photon anisotropic flow coefficients remains the same as those in nucleus-nucleus collisions, the splittings among the results using different emission rates are larger in 0-1\% p+Pb collisions at 5.02 TeV. The direct photon anisotropic flow coefficients in small collision systems show a strong sensitivity to the QGP photon emission rates.

\section{Concluding Remarks}\label{sec_4}
In this article, we review the evaluation of direct-photon production in heavy ion collisions from holography convoluted with the medium evolution and the inclusion of other sources \cite{Iatrakis:2016ugz}.
The agreement with experiments in spectra therein is improved compared with the previous study by using the pQCD rate. On the other hand, the deviation in flow is increased at low $p_T$ but decreased at high $p_T$. 

In small collision systems, where the experimental data of direct photons have not been available, holographic models lead to enhancements in both spectra and flow. The findings in \cite{Iatrakis:2016ugz} may emphasize the strong influence of thermal photons from the QGP phase on the direct-photon flow at high $p_T$, where hadronic contributions are highly suppressed. The enhancement of flow in this region stems from the amplification of the weight of late-time emission and the amplitude of thermal-photon emission in the QGP phase. As indicated in \cite{Iatrakis:2016ugz}, the "blue-shift" due to the increase of couplings could in general lead to the latter effect, whereas the former one is in fact model dependent.

On the contrary, the study may further {\em suggest that the hadronic contributions are responsible for the flow
at low $p_T$.} In contrast to the nucleus-nucleus collisions, the dominance of QGP photons could be more pronounced in small collision systems for larger $p_T$ window. Therefore, future measurements of direct-photon spectra and flow in small systems will be crucial to understand the electromagnetic property of the sQGP.

Acknowledgments:
The work of I.I. is part of the D-ITP consortium, a program of the Netherlands Organisation for Scientific Research (NWO) that is funded by the Dutch Ministry of Education, Culture and Science. The work of E. K. was partially supported by  European Union's Seventh Framework Programme under grant agreements (FP7-REGPOT-2012-2013-1) no 316165 and the Advanced ERC grant SM-grav, No 669288. D.Y. was supported by the RIKEN Foreign Postdoctoral Researcher program. C. S. was supported by the Natural Sciences and Engineering Research Council of Canada.

%
%


\end{document}